\documentclass[aps,reprint,onecolumn,amsmath,amssymb,amsfonts,floatfix,letterpaper]{revtex4-2}
\usepackage{scalerel}
\usepackage{ulem}
\usepackage{graphicx}
\usepackage[export]{adjustbox} 
\usepackage{times,float}
\usepackage[usenames,dvipsnames,svgnames]{xcolor}
\usepackage{multirow}
\usepackage{soul}

\usepackage[makeroom]{cancel}
\usepackage{dcolumn}
\usepackage{bm}
\usepackage{latexsym}
\usepackage{bm}
\usepackage{slashed}
\usepackage{svg}
\usepackage{stackengine}
\usepackage{color,soul}
\usepackage{fancyhdr}
\usepackage{epsfig}
\usepackage{pstricks}  
\usepackage[compat=1.1.0]{tikz-feynman}
\usepackage{tcolorbox}
\usepackage{epstopdf}
\usepackage{empheq}
\usepackage[colorlinks = true, linkcolor = blue, citecolor =unabred, urlcolor = unabblue]{hyperref}

\usepackage{amsbsy}

\usepackage{mathrsfs}
\usepackage{url}
\usepackage{empheq}
\usepackage{epsfig}

\newcommand{\bvec}[1]{\kern-.14em{\vec{\kern+.14em\bf#1}}} 
\newcommand{\mbf}[1]{\mathbf{#1}}

\newcommand{\bs}[1]{\boldsymbol{#1}}
\newcommand{\tongo}[1]{\hat{\mathbf{#1}}} 

\newcommand{\lsim}
  {\vcenter{\hbox{$\mskip6mu
    \hbox{$<$}\mkern-14.4mu\lower1.1ex\hbox{$\sim$}
  \mskip6mu$}}}                           
\newcommand{\gsim}
  {\vcenter{\hbox{$\mskip6mu
    \hbox{$>$}\mkern-14.4mu\lower1.1ex\hbox{$\sim$}
  \mskip6mu$}}}         

\def\sqr#1#2{{\vcenter{\vbox{\hrule height.#2pt
         \hbox{\vrule width.#2pt height#1pt \kern#1pt
         \vrule width.#2pt}
         \hrule height.#2pt}}}}

\def\lrpartial{\raise 1pt\hbox{$\stackrel\leftrightarrow\partial$}}

\def\tem{$\mathcal{T\!E\!M}$ }

\newcommand{\be}{\begin{equation}}
\newcommand{\ee}{\end{equation}}
\newcommand{\ba}{\begin{eqnarray}}
\newcommand{\ea}{\end{eqnarray}}

\definecolor{paleyellow}{rgb}{1,1,0.8}
\definecolor{unabblue}{rgb}{0.19,0.19,0.69}
\definecolor{unabred}{rgb}{0.67,0.11,0.18}

\begin{document}

\title{
New polarization rotation and exact TEM wave solutions in topological insulators
}
\author{Sebastián Filipini}
\email{s.filipiniparra@uandresbello.edu}
\affiliation{Universidad Andres Bello, Departamento de Ciencias Físicas, Facultad de Ciencias Exactas, Avenida República 220, Santiago, Chile}
\author{Mauro Cambiaso}
\email{mcambiaso@unab.cl}
\affiliation{Universidad Andres Bello, Departamento de Ciencias F\'isicas, Facultad de Ciencias Exactas, Avenida Rep\'ublica 220, Santiago, Chile}

\begin{abstract}
In the context of $\theta$ electrodynamics we find transverse electromagnetic wave solutions forbidden in Maxwell electrodynamics. Our results attest to new evidence of the topological magnetoelectric effect in topological insulators, resulting from a polarization rotation of an external electromagnetic field. Unlike Faraday and Kerr rotations, the effect does not rely on a longitudinal magnetic field, the reflected field, or birefringence. The rotation occurs due to transversal discontinuities of the topological magnetoelectric parameter in cylindrical geometries 
The dispersion relation is linear, and birefringence is absent. One solution behaves as an optical fiber confining exact transverse electromagnetic fields with omnidirectional reflectivity. These results may open new possibilities in optics and photonics by utilizing topological insulators to manipulate light.

\end{abstract}
\maketitle

\section{Introduction}
\label{EQ:Intro}
%
%
The topological magnetoelectric effect (TME) has been intensely sought after in recent decades as a definitive signal of quantum states of matter possessing topological order \cite{
haldane1984magnetic, kosterlitz1973ordering,
laughlin1981quantized,thouless1982quantized, kane_z2_PhysRevLett.95.146802, bernevig2006quantum_science.1133734, Moore_Balents_PhysRevB.75.121306,Fu_Kane_Mele_PhysRevLett.98.106803,Fu_Kane_PhysRevB.76.045302}. 
%
Topological insulators (TIs) are among the most well-known and studied cases presenting TME. These new quantum states can be found in heterostructures of elements like  Bi, Se, Te, Sb, and others \cite{chen2009experimental, hsieh_topological_2008, sato_DirarcCone_evidence_2010, xia_observation_2009, zhang_topological_2009
}. 
They exhibit conducting edge/surface states protected against disorder by time-reversal symmetry, with properties differing from those in the bulk of the material, which is gapped like conventional insulators \cite{hasan_colloquium_2010, RevModPhys.83.1057}.

Due to their microscopic structure, 
3D TIs  have unique electromagnetic (EM) responses that can be described macroscopically by the axionic $\theta$-term 
$\mathcal{L}_\theta = (\theta / 4 \pi)\mathbf{E}\cdot\mathbf{B}$ \cite{Wilczek:1987mv}.  In the context of TIs, $\theta =  \frac{\alpha}{\pi}\theta_{\textrm{TI}}$, where $\alpha$ is the fine-structure constant and $\theta_{\textrm{TI}}$ is called the topological magnetoelectric polarizability (TMEP). 
Its origin is quantum-mechanical and it encodes the microscopic properties that characterize TIs. 
This provides a correct description of the system if an appropriate time-reversal symmetry breaking perturbation is introduced to gap the surface states, which results in the material (in its bulk and at the surface) becoming an insulator. The surface, however, is a quantum Hall insulator rather than a normal one.  The latter can be achieved
by adding a magnetic perturbation (applied field and/or film coating) \cite{maciejko_topological_2010,qi_topological_2008}, or by using commensurate out- and in-plane antiferromagnetic or ferrimagnetic insulating thin films \cite{oroszlany_PhysRevB.86.195427}. %
As a result, $\theta_{\textrm{TI}}$ becomes quantized in odd-integer values of $\pi$ i.e., $\theta_{\textrm{TI}} = \pm (2n + 1)\pi$, where $n \in \mathbb{Z}$ and the sign is determined by the time-reversal symmetry breaking perturbation. Trivial insulators have $\theta_{\textrm{TI}} = 0$.
In this work, $\theta_{\textrm{TI}}$ will be taken  as a  constant parameter  characteristic of each medium. For brevity we will simply write $\theta$  
and we
shall refer to this theory as $\theta$-electrodynamics ($\theta$-ED) rather than axion electrodynamics. 
This model can also describe: general magnetoelectric media \cite{odell_1962electrodynamics, landau2013electrodynamics,shaposhnikov2023emergent}; metamaterials when $\theta$ is a purely complex function \cite{plum_metamaterial_2009}; and  Weyl semimetals when $\theta (\mathbf{x}, t) = 2 \left(\mathbf{b} \cdot \mathbf{x} - b_0 t \right)$ where $\mathbf{b}$ is the separation in momentum space between the Weyl nodes and $b_0$  their separation in energy \cite{vazifeh_electromagnetic_2013,martin-ruiz_electromagnetic_2019}. 
In this work, we will focus on TME signals stemming from the  EM response of TIs following closely the methodology of 
\cite{martin-ruiz_greens_2015, martin-ruiz_electro-_2016, martin-ruiz_magnetoelectric_2018} %
and also similar to what has been done, for example, to study Faraday rotation \cite{shuvaev_room_2013,shuvaev_terahertz_2013,crosse_electromagnetic_2015,crosse_theory_2017, wu_quantized_2016}, induced magnetic-monopole-like fields \cite{qi_monopole_2009}, and topologically induced effects in cavities and slab-waveguides \cite{
melo_topological_2016, campos_geometrically_2017}.
On the other hand, whenever a no-go theorem can be circumvented, a door into new theoretical and/or experimental possibilities is opened. In \cite{martin-ruiz_electro-_2016} it was shown  that $\theta$-boundary value problems can evade Earnshaw's theorem, which implies that transverse electromagnetic ($\mathcal{T\!E\!M}$) fields cannot propagate in media with less than two conductors %
%
\footnote{It is customary to call the transverse electromagnetic field as TEM. In this work, however, we introduce the typographical notation $\mathcal{T\!E\!M}$ so as to avoid confusion with the (very similar) TME or TMEP acronyms introduced before.}. 
Hence, as one of the most striking effects of $\theta$-ED is to modify the boundary conditions (BCs) that the fields must satisfy, in this letter we pursue this idea in systems that are heavily reliant on BCs to find novel $\mathcal{T\!E\!M}$ wave solutions that are impossible with topologically trivial material, and at the same time provide 
observable signatures of the elusive TME that are different from those previously reported in the literature. Our findings pave the way to new means of harnessing light with  possible applications in photonics that are yet to be discovered.

This manuscript is organized as follows. In Section \ref{SEC:Framework} we review the basics of $\theta$-electrodynamics. That is to say the field equations for Maxwell's Lagrangian appended with the axion term commented above, emphasizing how the $\theta$-term modifies the boundary conditions that the fields must satisfy at spatial surfaces where $\theta$ is discontinuous. The field equations are decomposed in longitudinal and transverse components as is customary for the study of field propagation in waveguides and/or optical fibers. In Section \ref{SEC:TEMwaves} we present the properties that the \tem  field possesses, namely, the relation defining the transversality condition, the general dispersion relation and the phase velocity. In this section we also introduce a  rotation of the plane of polarization of the EM field propagating transversely to $\bs \nabla \theta$ that is different to Faraday or Kerr rotations. In Section \ref{SEC:singlelayer} we present explicit solutions for the \tem fields inside and outside a single cylindrical TI with constant $\theta$ impinged upon by an external background EM field that serves as an asymptotic boundary condition. Most of the physics of this solution is analyzed through the field distribution as depicted in Fig. (\ref{FIG:DensityPlotEn1}). In Section \ref{SEC:polarizations} we comment on the role that different polarizations of the background EM field would have on the rotating effect of the TI and on the resulting spatial distribution of the EM field. Section \ref{SEC:severallayers} introduces the idea of considering several $\theta$-interfaces and the possibilities in terms of the possible configurations depending on the $\bs \nabla \theta$ at each surface. More specifically, in Section \ref{SEC:2layers} we analyze the case of two $\theta$-interfaces. This divides the whole space in three cylindrical regions: (a) $(0, R_1)$; (b) $(R_1, R_2)$; and (c) $(R_2, \infty)$. For the TMEP of each region we will choose them in the ``antiparallel'' configuration (see Fig. (\ref{FIG:Geo}.b)). That is when the gradient of $\theta$ at both  layers (and in the same angular direction) are  antiparallel, and to simplify the analysis, we will furthermore choose the inner and outer regions as topologically trivial, such that the geometry is basically that of a cylindrical TI shell of finite width.  In Section \ref{SEC:powers} we analyze the power transmitted in the different cylindrical regions defined by the $\theta$-interfaces and compare it with the power that would be transmitted through the same regions but without the TI.  In Section \ref{SEC:confinement} we elaborate criteria that allows us to speak of the confining capacity of the cylindrical TI shell on exact \tem fields that propagate in omnidirectional manner, acting as an optical fiber.
Finally in section \ref{SEC:conclus} we summarize our conclusions, provide some context as to the importance and relevance of finding \tem solutions besides the fact  of providing and alternative and different electromagnetic response of TI as evidence of the topological magnetoelectric effect, and elaborate on possible extensions and applications of these ideas. %
Throughout the paper, the equations of $\theta$-ED will be written in 
Gaussian units.  The coordinates $(\rho, \phi, z)$ are the cylindrical coordinates with $z$ in the direction of the wave propagation and of the cylindrical surfaces. $\rho$ and $\phi$ are the usual ones related to the Cartesian directions in Fig. (\ref{FIG:Geo}.a), i.e., $\bs \nabla \theta$ points in the radial direction $\bs{\tongo \rho}$, and $\bs{\tongo \phi}$ is perpendicular to the latter, in the anti-clockwise direction.

\section{Nondynamical $\theta$-electrodynamics}
\label{SEC:Framework}
\begin{figure}[h]
\stackinset{l}{2pt}{t}{3pt}{\Large{(a)}}{\includegraphics[width=0.21 \textwidth, frame]{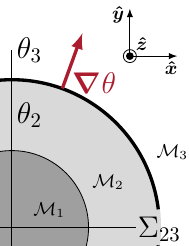}}
\hspace{20pt}
\stackinset{l}{2pt}{t}{3pt}{\Large{(b)}}{\includegraphics[width=0.21 \textwidth, frame]{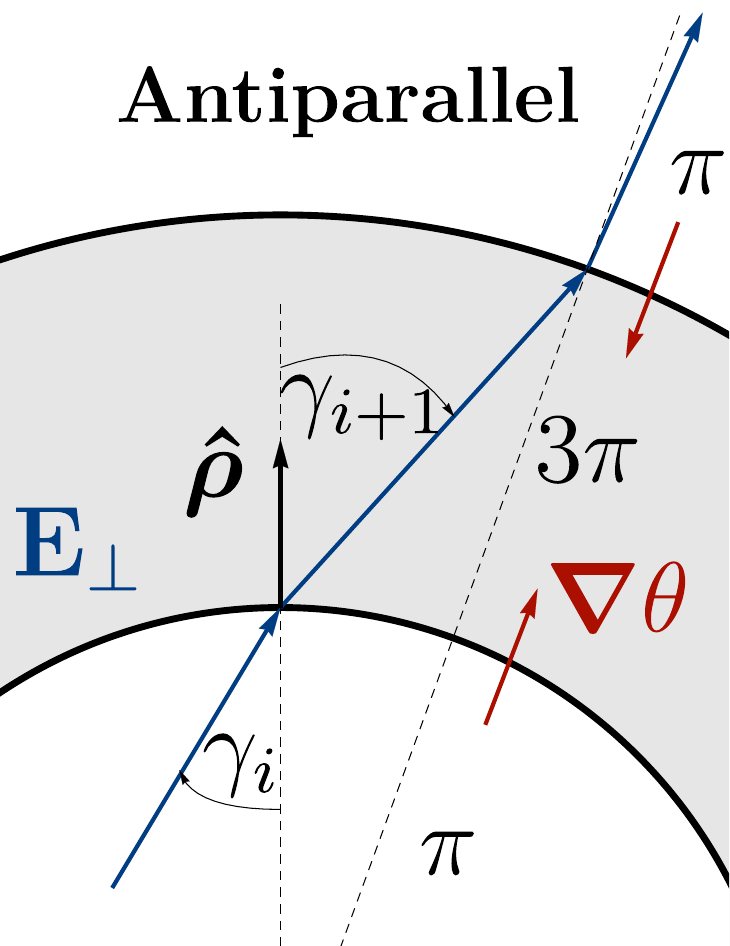}}
\hspace{20pt}
\stackinset{l}{2pt}{t}{3pt}{\Large{(c)}}{\includegraphics[width=0.21 \textwidth, frame]{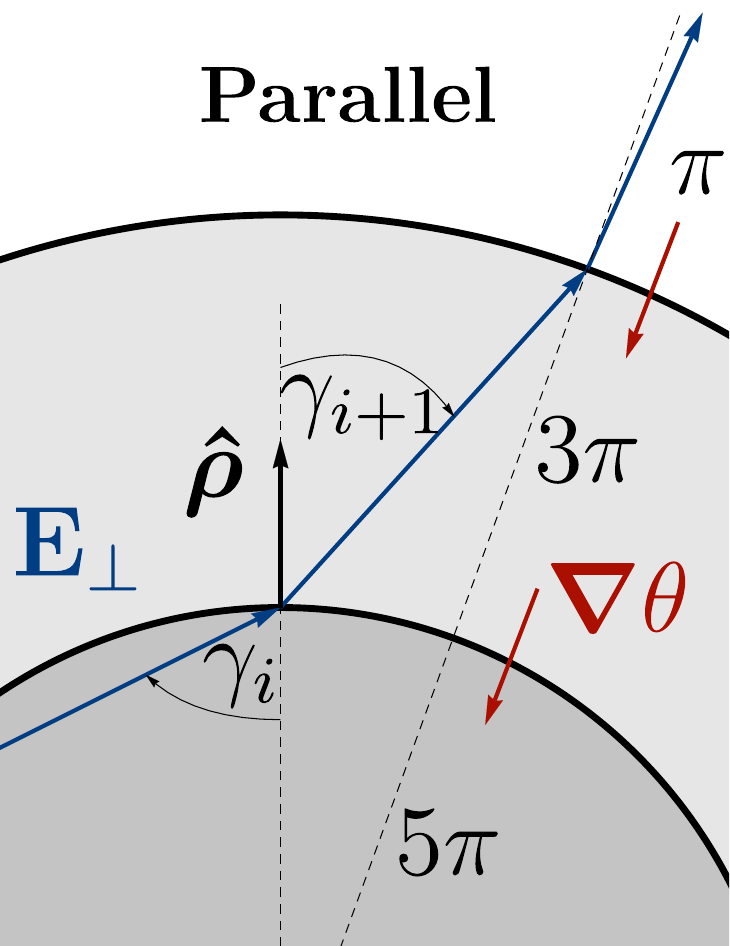}}
\caption{In (a) we show a generic cylindrical geometry. Only three media $\mathcal{M}_i$ are shown (each characterized by $\epsilon_i, \mu_i, \theta_i$). In this example $\theta_1 = 0$, so $\Sigma_{23}$, located at $R_2$ is the only $\theta$-interface and $\bs \nabla \theta = (\theta_3 - \theta_2)\delta(\rho - R_2)\boldsymbol{\hat\rho}$. In (b, c)   different $\bs \nabla \theta$ configurations are shown in which an EM wave propagates in the $\tongo z$ direction.  Antiparallel (b) and Parallel (c)  refer to the directions of $\bs \nabla\theta$ (red). The transverse $\mbf E_{\perp}$-fields are shown in blue.}
\label{FIG:Geo}
\end{figure}

In $\theta$-ED, the source-free equations do not change, but Gauss' and Amp\`ere-Maxwell do. The $\theta$-ED equations are:
\begin{align} 
    \boldsymbol{\nabla}\cdot(\epsilon\mathbf{E}) &=4\pi\rho-\boldsymbol{\nabla}\theta \cdot \mathbf{B},  \label{EQ:ThetaED_Gauss}\\  
    \boldsymbol{\nabla}\cdot\mathbf{B} &=0,  \label{EQ:ThetaED_monopole}\\  
    \boldsymbol{\nabla}\times  \mathbf{E} + \frac{1}{c} \frac{\partial \mathbf{B}}{\partial t}&=0. 
    \label{EQ:ThetaED_Faraday}\\
    \boldsymbol{\nabla}\times ( \mathbf{B}/ \mu)- \frac{1}{c} \frac{\partial(\epsilon\mathbf{E})}{\partial t}&= \frac{4\pi}{c} \mathbf{J}+\boldsymbol{\nabla}\theta \times \mathbf{E}+ \frac{1}{c} \dot{\theta} \,\mathbf{B}. 
    \label{EQ:ThetaED_AmpMax}
\end{align} 
The $\theta$-ED equations can be interpreted as if the field equations were not modified, but rather the constitutive relations were changed to: $\mathbf{D} \!=\! \epsilon \mathbf{E} + \theta \mathbf{B}$ and $\mathbf{H} \!=\! \mu^{-1}  \mathbf{B} - \theta \mathbf{E}$.
This makes manifest the role of $\theta$ as the culprit of the magnetoelectric effect, nevertheless, we will work out directly from Eqs. (\ref{EQ:ThetaED_Gauss}) - (\ref{EQ:ThetaED_AmpMax}) %
\footnote{%
Thus written Eqs. (\ref{EQ:ThetaED_Gauss}) and (\ref{EQ:ThetaED_AmpMax}) are not specific to TIs. They describe a wider class of magnetoelectric media, sometimes referred as Tellegen media \cite{tellegen1948gyrator}. 
If for whatever reasons one is not interested in the TME, one might as well forget about the TI altogether and focus on the EM response of general magnetoelectrics. For this reason we write $\theta$ rather than $\theta_{\textrm{TI}}$, but our calculations do consider the factor $\alpha/\pi$.
}.
We consider  $\theta(\mathbf{x},t )$ to be constant in time and throughout each medium, with constant and finite discontinuities at the interfaces between them, namely $\boldsymbol{\nabla} \theta = \tilde{\theta}_i \delta(f_{\Sigma_i}(\mathbf{x})) \tongo n_i$, where $\tilde{\theta}_{i}\equiv\theta_{i+1}-\theta_{i}$ and $\theta_{i}$ being the value of the TMEP in medium $i$. The interface is defined by $f_{\Sigma_i}(\mathbf{x})=0$ and $\tongo n_i$ is perpendicular to $\Sigma_i$ going from medium $i$ to medium $i+1$. The $\theta$ term does not modify the field equations in the bulk, but modifies the BCs as:
\begin{equation}
\Delta [\epsilon\mathbf{E}_{\perp }]|_{\Sigma }=-\tilde{\theta}\mathbf{B}_{\perp }|_{\Sigma } \quad \textrm{and} \quad   \Delta \left[\mu^{-1}\mathbf{B}_{\parallel} \right]|_{\Sigma }=\tilde{\theta}\mathbf{E}_{\parallel }|_{\Sigma } 
\label{EQ:BCs}
\end{equation}
where $\Delta [A]\equiv A_{i+1}-A_{i}$. Eqs. (\ref{EQ:BCs}) lead to different solutions both at the interfaces and in the bulk. In this work we will focus on cylindrical geometries with coaxial symmetry, an example of which is shown in Fig. (\ref{FIG:Geo}.a) and consider media separated by coaxial cylindrical surfaces $\Sigma$ and seek  monochromatic harmonic wave solutions for the EM fields with wave vector $ \mbf k = k \tongo z$, such that $\mathbf{E}(\mathbf{r},t)=\mathbf{E}(\mathbf{r}_{\perp})e^{i(kz-wt)}$ and similar for $\mathbf{B}(\mathbf{r},t)$. The axis are oriented such that $OZ$ coincide with the axis of the cylindrical surfaces.
As is common \cite{Jackson:1998nia}, we decompose vectors in directions longitudinal and transverse to the direction of propagation
and the vacuum field equations in each medium  read:
\begin{eqnarray}
\epsilon\boldsymbol{\nabla}_{\perp} \cdot  \mathbf{E}_{\perp}+ [ \tilde \theta \mbf{B}_{\perp}]_\rho&=&-ik\epsilon E_{z}\,, \qquad \qquad 
\boldsymbol{\nabla}_{\perp}  \cdot \mathbf{B}_{\perp}=-ik B_{z}, \,
\label{EQ:waveguide_eq1}\\
 ik\mathbf{E}_{\perp}+ik_{0}\, \tongo{z}\times \mathbf{B}_{\perp} &=&  \boldsymbol{\nabla}_{\perp}E_{z}\,, \qquad \qquad
 \tongo z \cdot (\boldsymbol{\nabla}_{\perp}  \times \mathbf{E}_{\perp})=ik_{0}B_{z} 
 \label{EQ:waveguide_eq2}\\
 ik \mathbf{B}_{\perp} - i \epsilon \mu k_0 \,  \tongo{z}\times  \mathbf{E}_{\perp} & = & \boldsymbol{\nabla}_{\perp}B_{z}-\mu [\tilde \theta E_{z}  \boldsymbol{\hat \rho}]
 \label{EQ:waveguide_eq3}\\
 \tongo z \cdot (\boldsymbol{\nabla}_{\perp}  \times \mathbf{B}_{\perp})  -\mu  [\tilde \theta \mathbf{E}_{\perp}]_{\phi} & = & -i\epsilon \mu k_{0}E_{z} 
  \label{EQ:waveguide_eq4}
\end{eqnarray}
where $k_0 \equiv \omega/c$ and in our case $\partial_z \theta$ and $\dot \theta$ terms vanish. Since  $\boldsymbol{\nabla} \theta$ has support at the given interfaces only we have put e.g.,  $\boldsymbol{\nabla}_{\perp} \theta \cdot \mathbf{B}_{\perp} =( \tilde \theta \mbf{B}_{\perp})_\rho|_\Sigma \equiv [\tilde \theta \mbf{B}_{\perp}]_\rho$.

\section{TEM  wave solutions and rotation of the plane of polariazation as a topological magnetoelectric signature}
\label{SEC:TEMwaves}

In \cite{martin-ruiz_electro-_2016} it was  reported for the first time that, in the context of TIs, $\bs \nabla \theta$ could evade the restrictions imposed by Earnshaw's theorem. In this work we present explicit examples of field solutions  now made available to us by the modifications introduced by $\theta$-ED that,  at the same time, interact with the TIs in a way to produce a novel signature of the TME.

The field equations (\ref{EQ:waveguide_eq1})-(\ref{EQ:waveguide_eq4}) admit self-consistent non-trivial solutions for the transverse components of  the electric and magnetic fields, i.e., $\mbf E_\perp \neq 0$ and $\mbf B_\perp \neq 0$ with $E_z = B_z = 0$ simultaneously provided: (a) $\mathbf{B}_{\perp}$ is transverse to $\mathbf{E}_{\perp}$:
\begin{eqnarray}
\label{EQ:BTEM}
\mathbf{B}_{\perp}&=\sqrt{\epsilon \mu}\, \tongo{z}\times\mathbf{E}_{\perp}.
\end{eqnarray}
(b) The  dispersion relations is:
\begin{equation}
c^2\,k^2 = \omega^{2} \, \mu\epsilon ,
\label{EQ:DispRel}
\end{equation}
where $\partial_z \theta$ and $\dot \theta$ terms, that are present in the general case, vanish, since we are working with the restricted $\theta$ for which these terms are taken to be zero.
(c) correspondingly, the  phase velocity $v_{p}(k)\equiv  \omega(k)/k$ is:
\begin{align}
   v_{p}(k)&=\frac{c}{\sqrt{\mu\epsilon}}, 
\end{align}
(d) and also, the optical properties across adjacent media  satisfies %
\footnote{This condition is not specific to axion electrodynamics. Actually if the usual hollow cylindrical coaxial waveguide were filled with different bilayered coaxial media, the same condition holds for $\mathcal{T\!E\!M}$ modes to exist. } %
$\epsilon_{i+1}  \mu_{i+1} = \epsilon_i \mu _i$. Given Eqs. (\ref{EQ:BTEM}) and (\ref{EQ:DispRel}) and our restriction for the TMEP, that is $\partial_z \theta = 0 = \dot \theta$, we observe that the fields propagate with continuous wavenumber, without birefringence, and with a dispersion relation as in a dispersion-free medium %
\footnote{By this we mean, no dispersion associated to a constant $\theta$ and also barring frequency dependence of $\theta$ \cite{ahn_theory_2022}, and of the permittivity $\epsilon$, which will
naturally cause dispersion.}
: $k = |\mbf k| = k_{0} \sqrt{\mu \epsilon}$ , i.e., as free  $\mathcal{T\!E\!M}$ waves in an $(\epsilon,\mu, \theta)$-medium. In what follows we will drop the subscript $\perp$ of the fields. 
The BCs imposed by the $\theta$-interface produce a discontinuity of $\mbf E$ field across the interface that results in a rotation of the polarization of the field that is solely due to $\bs \nabla \theta$ across the interface. This situation is depicted in Figs. (\ref{FIG:Geo}.b, c). Note the importance of the sign of $\tilde \theta$ and that  only the radial component of $\mbf E$ is discontinuous. At any given point of the $\theta$-interface, the directions of the $\mbf E$  satisfy:
\begin{equation}
\label{EQ:refracangle}
\tan \gamma_{i+1}= \tan \gamma_{i} \frac{1}{\left(1+2 Z_\theta \tan \gamma_i \right)},
\end{equation}
where $Z= \sqrt{\mu/\epsilon}$ is the impedance, $Z_{\theta} \equiv \tilde \theta Z/2$,  and $\gamma_i$,  $\gamma_{i+1}$ and are the angles between the normal to the $i$-th interface and $\mbf E$ on either side of it. Faraday and Kerr rotation effects, have indeed been predicted in the context of TIs as signals of the TME \cite{ahn_theory_2022, shuvaev_giant_2011, shuvaev_room_2013, shuvaev_universal_2022, crosse_theory_2017, crosse_optical_2016, wu_quantized_2016, 
tse_giant_2010,yang_transmission_2018}. The rotation we find here is also an interesting signature of the TME, but owing to the \tem nature of our solution, it differs radically from the latter. Contrary to Faraday rotation, this one is not generated by a component of the $\mbf B$-field along the direction of propagation (because the fields are transverse). Neither is it due to  birefringence, as we have $ck = \omega \sqrt{\mu \epsilon}$, nor is it a property of the polarization of the reflected field. This is a novel prediction of the EM response of TIs leading to a new way to observe the TME, and  it is a consequence of  exact \tem wave solutions that had not been exploited up until now and are certainly very intensively sought for  \cite{ibanescu_all-dielectric_2000, nordebo_dispersion_2014,catrysse_transverse_2011,shvedov_topological_2017, zhou_surface_2023}.  If one considers a geometry with several coaxial cylindrical layers, the cumulative effect is different for the parallel and antiparallel configurations of Fig. (\ref{FIG:Geo}.b,c), because the effect is sensitive to $\tilde \theta$. We now pass to analyze particular cylindrical configurations. For the sake of separating the $\theta$-effect from other possible optical effects, in the remaining we will consider $Z=1$.

\section{One cylindrical $\theta$-interface}
\label{SEC:singlelayer}
Consider an infinitely long TI cylinder of radius $R$, characterized by $\theta$  in a homogeneous medium, both with the same $Z$.  We seek solutions such that asymptotically away from the TI cylinder, the EM field be a plane wave with linear polarization (LP), say, in the $\tongo y$-direction, i.e., $\lim_{\rho\to\infty}\mathbf{E}(\rho,\phi,z,t)=E_{0}\,e^{i(kz-\omega t)}\, \tongo{y}$.

\begin{figure}[h]
\begin{minipage}{0.4\columnwidth}
\stackinset{l}{25pt}{t}{15pt}{\Large{(a)}}{\includegraphics[width=0.85\textwidth]{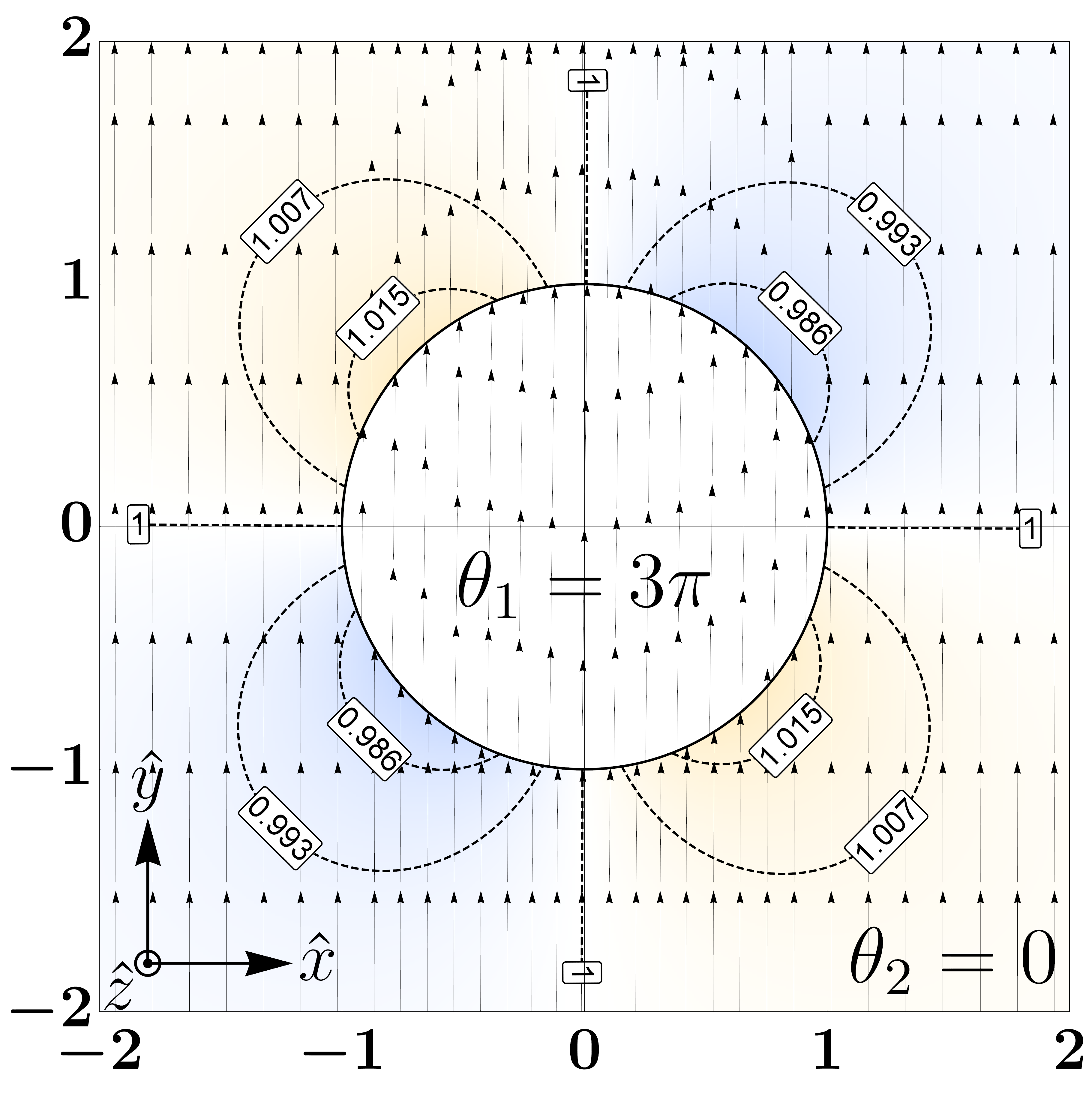}}
%
\vspace{6mm}

%
\stackinset{l}{25pt}{t}{15pt}{\Large{(b)}}{\includegraphics[width=0.85\textwidth]{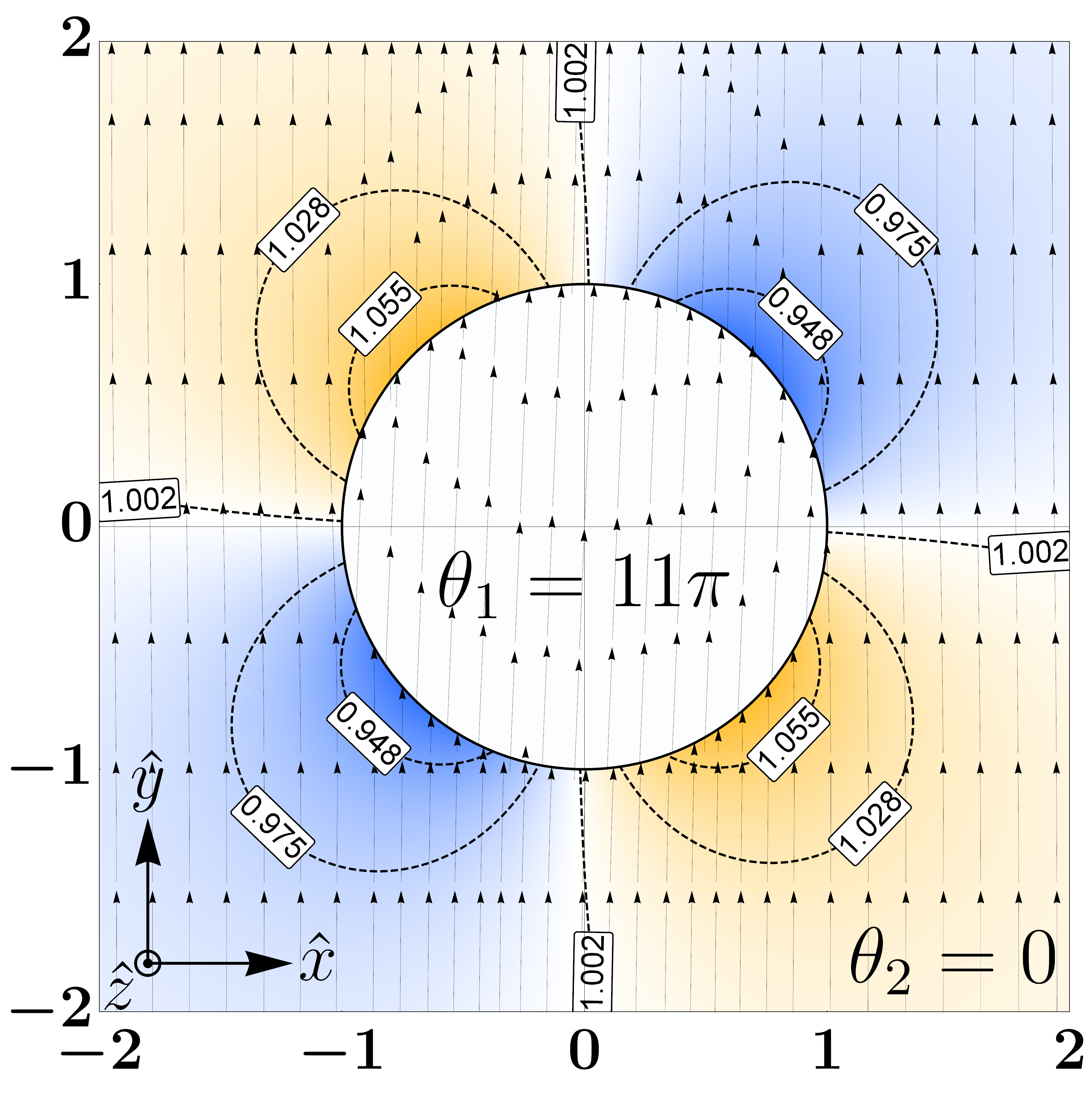}}
\end{minipage}
\begin{minipage}{0.4\columnwidth}
\stackinset{l}{25pt}{t}{15pt}{\Large{(c)}}{\includegraphics[width=0.85\textwidth]{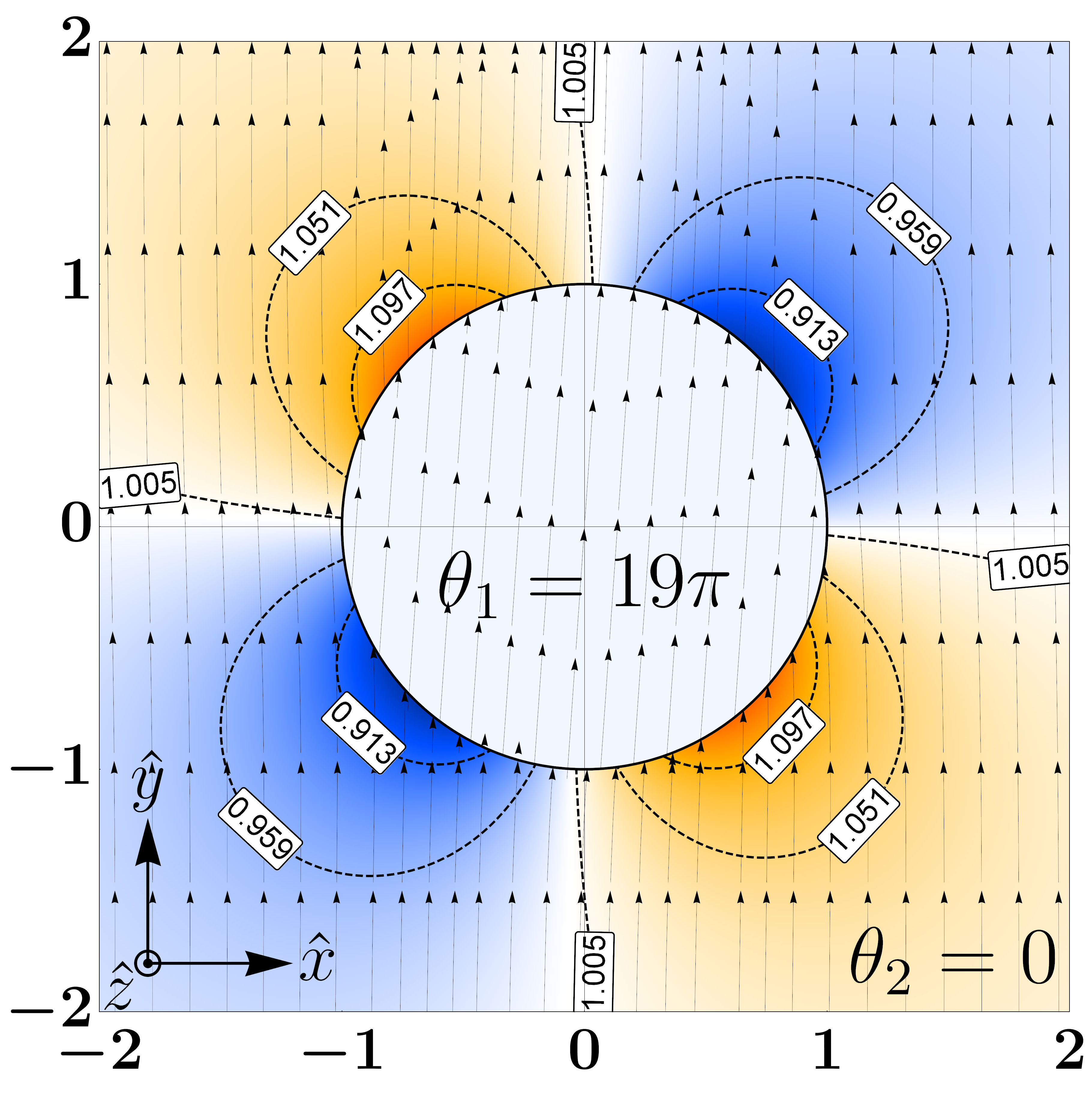}}
%
\vspace{6mm}

%
\stackinset{l}{25pt}{t}{15pt}{\Large{(d)}}{\includegraphics[width=0.85\textwidth]{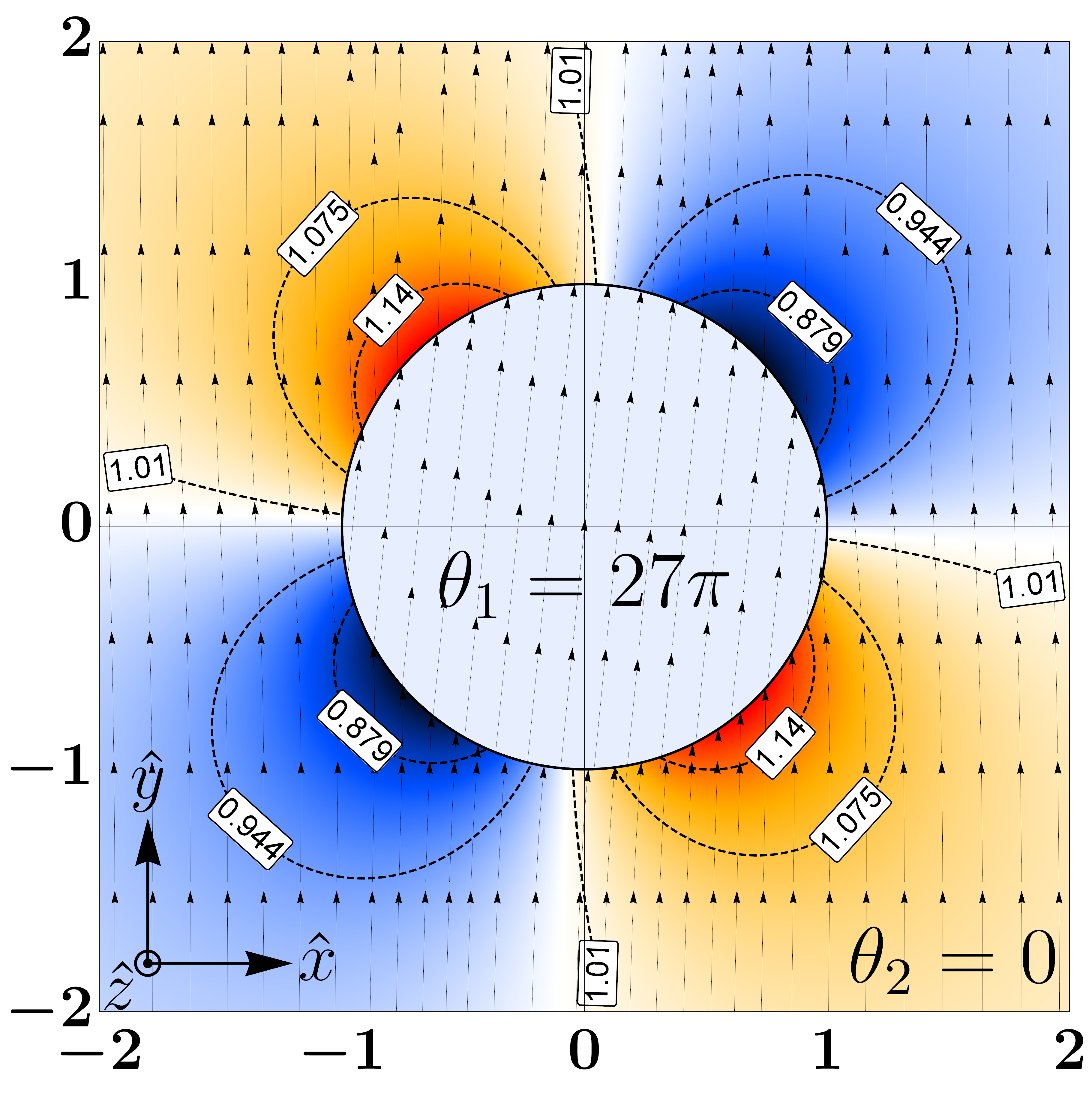}}
\end{minipage}
\begin{minipage}{0.08\columnwidth}
\includegraphics[scale=0.07]{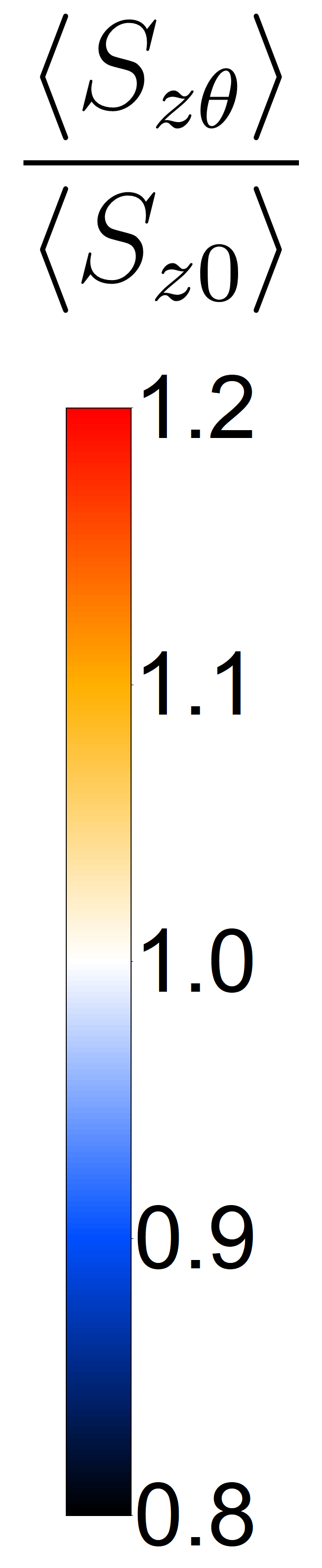}
\end{minipage}
\caption{%
The color map is a density plot of the temporal average of the total  Poynting vector in the interior and exterior regions relative to that of the background EM field, $ \langle S_{z0}  \rangle=c E_{0}^{2}/8\pi Z$. The streamlines are the total $\mbf E$ field-lines. The contour plots show contours of constant relative Poynting vector. Panels (a-d) correspond to $Z=1$ and $\theta =3\pi, 11\pi, 19\pi, 27\pi$, respectively. 
}
\label{FIG:DensityPlotEn1} 
\end{figure}

The \tem fields that solve Eqs. (\ref{EQ:waveguide_eq1})-(\ref{EQ:waveguide_eq4}) and satisfy the BCs of Eq. (\ref{EQ:BCs}), in each media $\mathcal{M}_i$, for $i=1,2$  are $\mathbf{E}_{i}=E_{0}\tongo{y}+E_{0}\mathbf{E}_{i}^{\theta}$, where:
\begin{align}
\mathbf{E}_{1}^{\theta}&\!=\!-\kappa (\tongo{x}+ Z_{\theta}\tongo{y})\,,
\label{EQ:Ethetatot1}\\\
\mathbf{E}_{2}^{\theta}&\!=\!\kappa \ell^2 [ ( Z_\theta  \sin\phi+\cos\phi)\boldsymbol{\hat{\rho}}+\!(\sin\phi-\!Z_\theta  \cos\phi)\boldsymbol{\hat{\phi}} ],
\label{EQ:Ethetatot2}
\end{align}
with $\kappa=Z_{\theta}/(1+Z_{\theta}^{2})$ and $\ell=R/\rho$. As $\mbf B$ is given by Eq. (\ref{EQ:BTEM}), in the sequel we will mostly refer to the electric field. If $\tilde \theta = 0$ there is no interface whatsoever and the interior and exterior solutions are identical to $E_{0}\,e^{i(kz-\omega t)}\, \tongo{y}$ as they must. If $\tilde \theta \neq 0$ and $E_0\neq 0$ the total field is non-trivial, but if $E_0 = 0$ there is no solution at all.  Therefore, our solution is reliant on the ``background'' field (we prefer to call this background rather than external so as not to confuse it with the field outside or exterior to the TI). The claim is not that this $\mathcal{T\!E\!M}$ solution exists solely due to the $\theta$-interface, but rather that due to it,  a solution exists in all space that cannot be otherwise obtained with, for example, all-dielectric materials, and it acquires new and non-trivial features that are attributable to $\theta$ alone that can lead to new observables signatures of TMEP.
In Fig. (\ref{FIG:DensityPlotEn1}) we show streamlines of the $\mbf E$-field. The density plots represent the spatial distribution of the temporal average of the total Poynting vector, relative to that of the background field, $ \langle S_{z0}  \rangle=c E_{0}^{2}/8\pi Z$. By total we mean the contribution to the Poynting coming from the total EM field, that is the superposition of the background EM field and the $\theta$-induced one. Observe that due to the \tem nature of the EM field, the Poynting vector only has longitudinal $z$ component. Inside the cylindrical TI, the electric field is  uniform, and, due to the  rotation of Eq. (\ref{EQ:refracangle}), the plane of polarization rotates by a fixed amount. As expected, the magnitude of the effect is minute, however, for feasible values of the TMEP parameter the rotation of the polarization plane is measurable with present day techniques. This rotation is given by $ \cos \varphi_{\textrm{int}} = \tongo{\mbf E}_{1} \cdot \tongo y = (1 + Z^2_\theta )^{-1/2}$. For example, for $\theta_{\textrm{TI}}= 3 \pi, 11 \pi, 19 \pi$ and $27 \pi$ respectively, this effect results in a rotation of the polarization plane by $0.63, 2.30, 3.97$ and $5.63$ degrees  respectively. This rotation  is entirely due to the TMEP of the cylindrical TI, it is of a completely different nature than Faraday or Kerr rotation,  and thus, it is yet another application of $\theta$-ED that provides an alternative method to measure a signal of the TME.
The temporal averages of the total relative Poynting vector in each region are given by:
\begin{align}
\langle S^\theta_{z1} \rangle/ \langle S_{z0} \rangle\!&=1-\kappa  Z_\theta, \label{EQ:InnerPoynting1layer}\\
\langle S^\theta_{z2} \rangle/ \langle S_{z0} \rangle\!&=\!1\!+\kappa\! \left [Z_\theta \ell^4\!+\! 2 \ell^2  (\sin 2 \phi-Z_\theta\cos 2 \phi)\right]\!. \label{EQ:OuterPoynting1layer}
\end{align}
Away from the TI's surface, the power per unit area varies as: (a) an anisotropic term that goes as $\rho^{-2}$, and (b) also by an isotropic term that goes as $\rho^{-4}$. The patterns of the relative Poynting vector ($\mathcal{S}_z^\theta(\rho,\phi) \equiv \langle S_{z\theta} \rangle / \langle S_{z0} \rangle$) in Fig (\ref{FIG:DensityPlotEn1}) reveal other interesting features. 
It appears that $\mathcal{S}_z^\theta(\rho,\phi)$ is distributed with a sort of quadrupolar structure in the $XY$-plane being slightly amplified/diminished alternately in each quadrant. Also, it seems its value at the $y=0$ and $x=0$ planes were equal to 1. %
This, however, is only an illusion due to the smallness of $\theta$-effect. The contour plots in Fig. (\ref{FIG:DensityPlotEn1}) show lines of constant  $\mathcal{S}_z^\theta(\rho,\phi)$ and illustrate this. The physics is neither left-right nor up-down symmetric, and it all is due to the interplay between the BCs and the fact that the background EM wave is comprised of $\mbf E$- and $\mbf B$-field vectors in the $+ \tongo y$ and $-\tongo x$ directions respectively. A correct interpretation of the patterns of $\mathcal{S}_z^\theta(\rho,\phi)$ resides on the symmetries of Eq. (\ref{EQ:OuterPoynting1layer}). One immediately sees that $\mathcal{S}_z^\theta(\rho,\phi) = \mathcal{S}_z^\theta(\rho,\phi + n \pi)$ for $n=1,2, \dots$ and also $\mathcal{S}_z^\theta(\rho,\phi)-\mathcal{S}_z^\theta(\rho,-\phi)= 4 \kappa \ell^2 \sin 2 \phi = \mathcal{S}_z^\theta(\rho,\phi)-\mathcal{S}_z^\theta(\rho,\pi -\phi)$. Therefore, though for fixed $\rho$ the relative Poynting is equal at antipodal points in $\phi$, it is not left-right nor up-down symmetric. Furthermore, the relative Poynting, as a function of $\phi$ has maxima and minima defined by the directions $\phi_{\pm} = \arctan (Z_\theta \pm \sqrt{1+Z^2_
\theta})$ respectively. 
These directions correspond to the lines (not drawn) on Fig. (\ref{FIG:DensityPlotEn1}) of extremal intensities. Given the asymmetries above, evaluated at the boundary, $|\mathcal{S}_z^\theta(R,\phi_+) - 1| > |\mathcal{S}_z^\theta(R,\phi_-) - 1|$, however, with respect to the directions of minimum and maximum value, the relative Poynting is indeed symmetric, namely, $\mathcal{S}_z^\theta(\rho,\phi_\pm  + \alpha) = \mathcal{S}_z^\theta(\rho,\phi_\pm  - \alpha)$. 
This asymmetric field distribution can be understood  self-consistently, order-by-order in $\theta$, in terms of  the induced topological surface charge densities. The jump in $\theta$ across the boundary times the $\mbf B$-field normal to the cylinder generates a discontinuity of the $\mbf E$-field that acts as a topological surface charge density $\sigma_\theta(\Sigma) = -\frac{1}{4\pi} (\tilde \theta  \mathbf{B} \cdot \bs{\hat{\rho}})|_{\Sigma}$. Along with it, there is an induced (topological) surface current density $\mathbf{K}_\theta(\Sigma)\! =\! \frac{c}{4\pi} (\tilde \theta \bs{\hat{\rho}} \times \mathbf{E})|_\Sigma$ %
\footnote{Since the surface charge and current densities $\sigma_{\theta}$ and $\mathbf{K}_{\theta}$ do satisfy a continuity equation, we will mostly refer to  $\sigma_{\theta}$. We note in passing that $\mbf K_{\theta} \cdot \mbf E =0$, hence these induced charges and currents produce no Ohmic losses.}. 
The total electric field $\mbf{E}^\theta_{1,2}$ (and the corresponding  $\mbf{B}^\theta_{1,2}$) can be understood as an infinite superposition of the fields induced by these topological surface charge densities and currents. The infinite sum, in fact converges and lead precisely to the fields in Eq. (\ref{EQ:Ethetatot1},\ref{EQ:Ethetatot2}). %
Further details in \cite{underprep}.

\subsection{The role of the polarization of the background field}
\label{SEC:polarizations} 
If the background field has right-handed or left-handed circular polarizations (RCP/LCP) the amount of the polarization rotation inside the TI is the same and in each case it is in the same sense the background field rotates.
At a given time and for appropriately chosen initial conditions (or phase) the total field structure for the CP background field and the patterns of the Poynting vector are the same as for the LP. The temporal averages differ considerably, though. For the CP background field, the pattern of the external Poynting is  isotropic only with a $\sim \tilde \theta^2\rho^{-4}$  dependence %
\footnote{For elliptical polarization similar conclusions can be drawn. Subtle differences  will be dealt with elsewhere \cite{underprep}.}. 
To understand this, we realize  that the Poynting has an interaction term $2 E_0 \tongo y 
\cdot E_0 \mbf{E}^{\theta \ast}$ in either regions interior and exterior to the TI. Going back to our discussion of the induced topological polarization charges, for a CP background field, these $\sigma_\theta$ will also tend to redistribute following the direction of the $\mbf E$ field,  which is rotating itself so there is no misalignment between the background field and that produced  by the induced charges,  thus they remain orthogonal to each other at all times.

\section{Several coaxial cylindrical $\theta$-interfaces}
\label{SEC:severallayers}
The precise form of the repeated effects with several coaxial cylindrical $\theta$-interfaces depend on the different radii at which the $\theta$-interfaces lie, on  $\bs \nabla \theta$ at each layer and possibly on the polarization of the background EM field. With respect to the directions of $\bs \nabla \theta$ there as several possible configurations. 
For simplicity we analyze the case of two $\theta$-interfaces and focus on  the  antiparallel  configuration, as depicted in Fig.  (\ref{FIG:Geo}.b), respectively. The study of more $\theta$-interfaces is left for elsewhere \cite{underprep}.

\subsection{Two coaxial $\theta$-interfaces in antiparallel configuration}
\label{SEC:2layers}

Consider now two coaxial cylindrical $\theta$-interfaces in antiparallel configuration with the same background EM field as above. The geometry is as in Fig. (\ref{FIG:Geo}.a),  with $\theta = \theta_2 \neq 0$ for $R_1 \leq \rho < R_2$ and zero elsewhere. Region 1  defined by $0 \leq \rho < R_1$, is the internal vacuum; region 2, defined by $R_1 \leq \rho < R_2$, is the TI's bulk; and region 3,  defined by $R_2 \leq \rho$, is the external vacuum. The ratio $\chi = R_1/R_2$  will be useful. 
In regions $i=1,2,3$ the total electric field is actually $\mathcal{T\!E\!M}$ and can be written as $\mathbf{E}_{i}=E_0 \tongo{y}+ E_0 \mathbf{E}_{i}^{\theta}$ and  the $\theta$ contributions are: 
\begin{align}
    \mathbf{E}_{1}^{\theta} &=-\Theta_\chi Y\theta_2\,  \tongo{y}\,, \label{EQ:E1theta}\\
    \mathbf{E}_{2}^{\theta} &= \Theta_{\chi}\Bigl[2\frac{R_1^{2}}{\rho^{2}}(\cos\phi\boldsymbol{\hat{\rho}}+\sin\phi\boldsymbol{\hat{\phi}})+2\tongo{x}-Y\theta_{2}\tongo{y}\Bigr]\,, \label{EQ:E2theta}\\
   \mathbf{E}_{3}^{\theta}&=\Theta_\chi Y\frac{R_2^2\, }{\rho ^2} \Bigl[ \sin \phi (\theta_2 \boldsymbol{\hat{\rho}} - 2\boldsymbol{\hat{\phi}}) -\cos \phi(\boldsymbol{\hat{\rho}}+\theta_2 \boldsymbol{\hat{\phi}} )\Bigr],
   \label{EQ:E3theta}
\end{align}
where $Y=Z(1-\chi^2)$ and $\Theta_\chi =  Z\theta _2/ (4 + Y Z \theta_2^2)$.  Despite the value of $\theta_2$,  the field in region 1 is uniform, with the  same polarization as the asymptotic background field and  $\chi$  determines how much is the intensity diminished in region 1. For $\chi=1$ (i.e.,  $R_1=R_2$) there is no additional $\theta$-contribution to the total field in region 1, as it should. 
In Figs. (\ref{FIG:DensitiyPlotAPwaveguide} a, b) we show the density plots of the temporal average of the Poynting vector relative to the background, and $\mbf{E}$-field streamlines, corresponding to Eqs. (\ref{EQ:E1theta}, \ref{EQ:E2theta}, \ref{EQ:E3theta}), for $Z=1$ and  $\theta_2 = 27 \pi$. For smaller $\theta$ the same effects arise, but fainter. In (a) $\chi_a=0.45$ and in (b) $\chi_b = 0.82$, respectively. In either case, in TI's bulk  the field is similar in its asymmetric quadrupolar-like  distribution, as for one $\theta$-layer, but, it is inverted with respect to the exterior region. The successive application of the BCs and the geometry give rise to a distribution of induced topological charges that generates the corresponding total electric field. 

\subsection{Transmitted power in each region as  a function of $\theta$ and the geometry of the system } \label{SEC:powers}

In Fig. (\ref{FIG:DensitiyPlotAPwaveguide}.c), for different values of $\theta_2$, we compare the power transmitted in region 1:
\begin{equation}
   P_{1}^{\theta_2}(\chi) = \int_{0}^{R_1} \langle S_{z\theta} \rangle ds\,,
\end{equation}
to the power transmitted in that same region by the background field: $P_1^{\theta_2=0}$  (in solid black line). For fixed $\chi$, we see that $P_1$ is smaller for higher values of $\theta$ and, for a given $\theta$, $P_1$ scales with $R_1$ (which is rather trivial as the bigger/smaller $R_1$ the bigger/smaller the area pierced by the Poynting). 
The differences $\Delta P_i (\chi) \equiv P_i^{\theta_2}- P_i^{\theta_2=0}$ for $i=1,2$ serve as a means to quantify the ``confining'' properties of the system. The quantities $P^{\theta, 0}_2(\chi)$ correspond to the transmitted power by the EM field through the region $R_1 \leq \rho \leq R_2$ with TI ($\theta$) and without (0), respectively, namely:
\begin{equation}
   P_{2}^{\theta_2}(\chi) = \int_{R_1}^{R_2} \langle S_{z\theta} \rangle ds\,.
\end{equation}
Their $\chi$ dependence allows to find optimal configurations.  For example, for any given $\theta$, there is a critical $\chi = \chi_{1M}$ that minimizes $\Delta P_1$. The inset to Fig. (\ref{FIG:DensitiyPlotAPwaveguide}.c) shows this difference and the values $\chi_a$ and $\chi_{1M}$ for $\theta_2=27 \pi$.
Rather surprisingly, in the $R_1 \to R_2$ limit, a radial electric anisotropic field residing only at $\rho=R_2$ remains while in regions 1 and 3 the total electric field is exactly equal to the background field.
\subsection{Geometry optimization and confinement of the \tem field inside the TI}
\label{SEC:confinement}

\begin{figure}[h!]
\hspace{-0.5cm}
\stackinset{l}{25pt}{t}{15pt}{\Large{(a)}}{\includegraphics[width=0.38\textwidth]{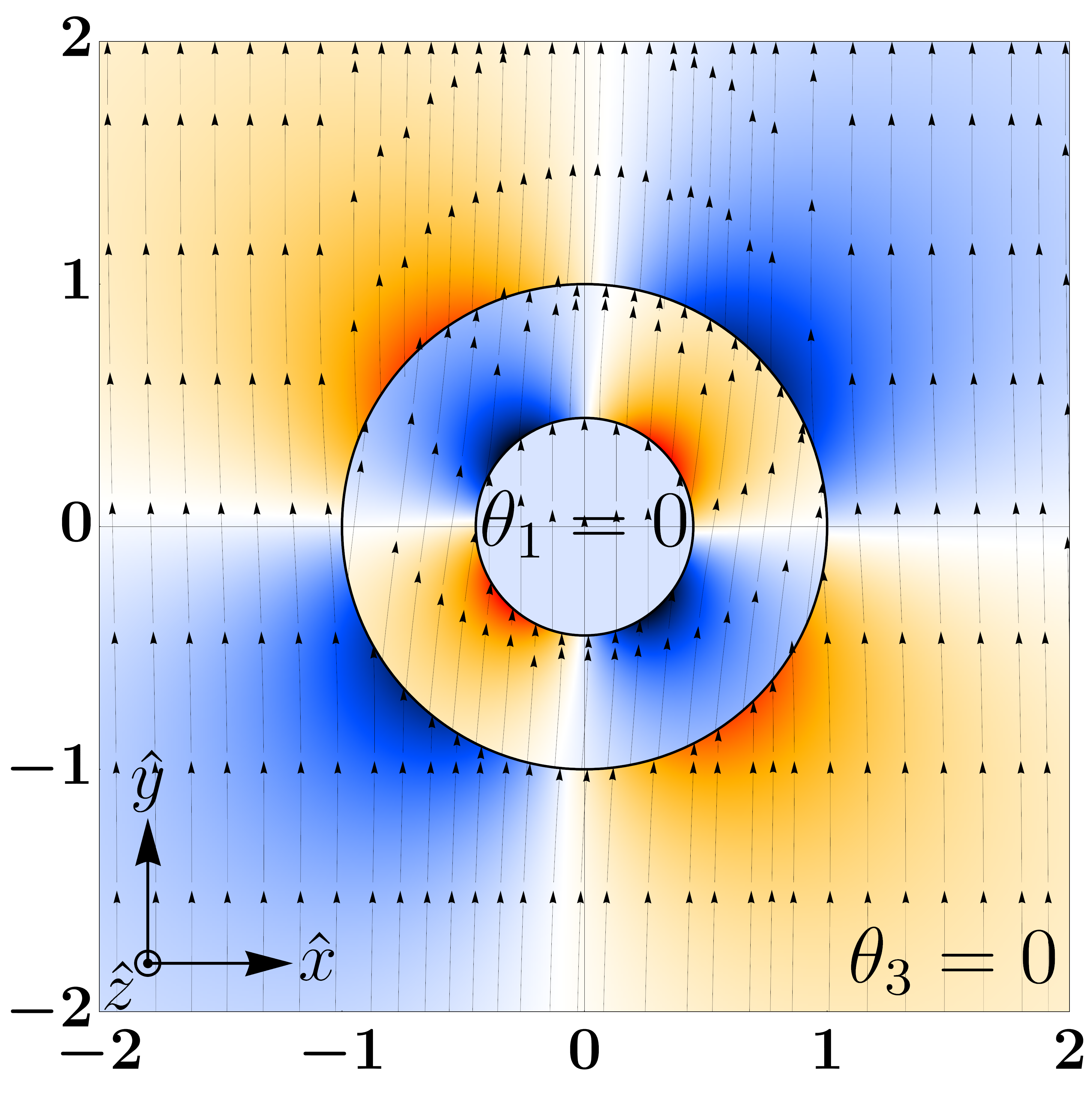}}
%
\hspace{1cm}
%
\stackinset{l}{25pt}{t}{15pt}{\Large{(b)}}{\includegraphics[width=0.455\textwidth]{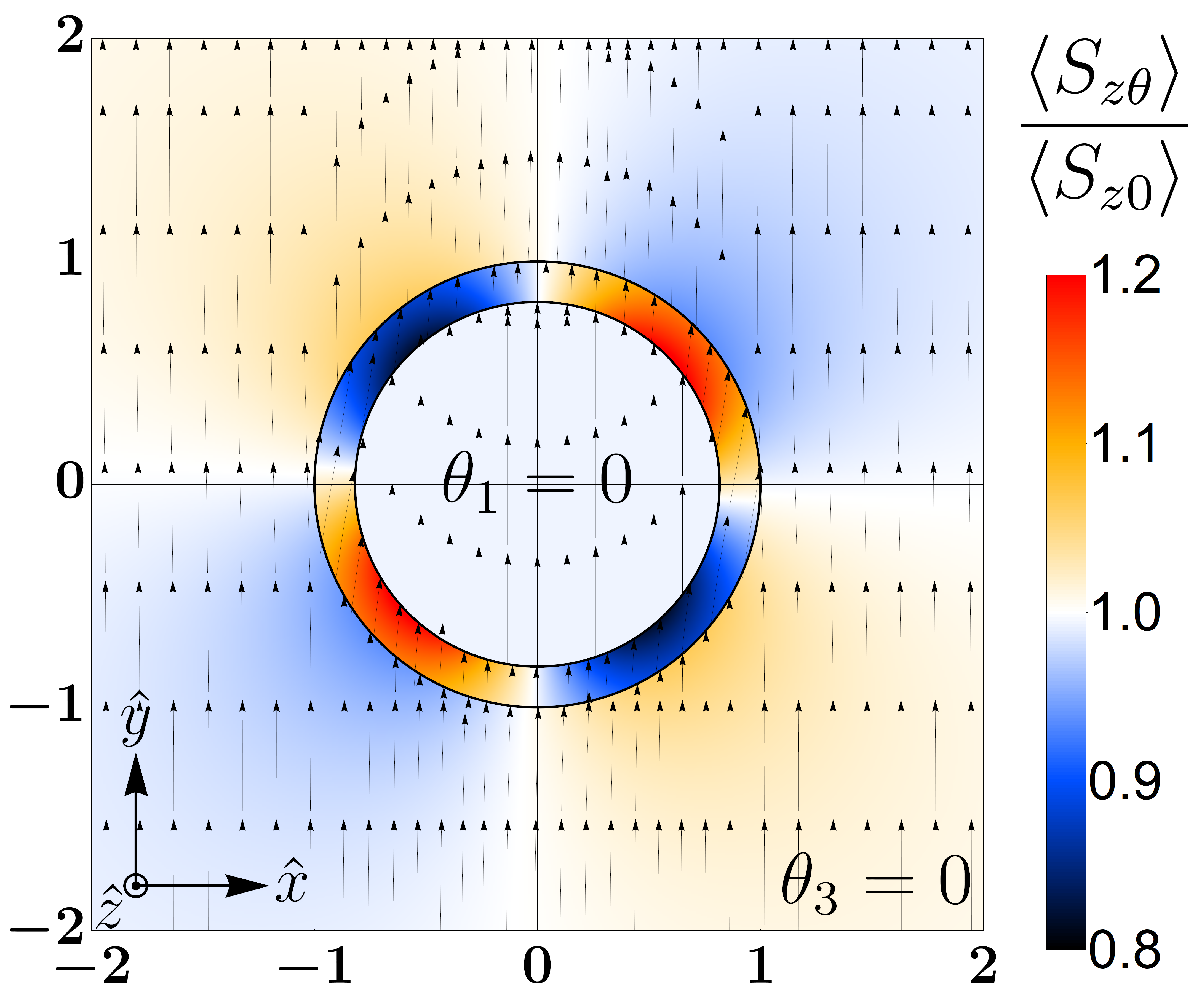}}

\vspace{9mm}

\hspace{-1.1cm}
\stackinset{l}{25pt}{t}{12pt}{\Large{(c)}}{\includegraphics[width=0.4\textwidth]{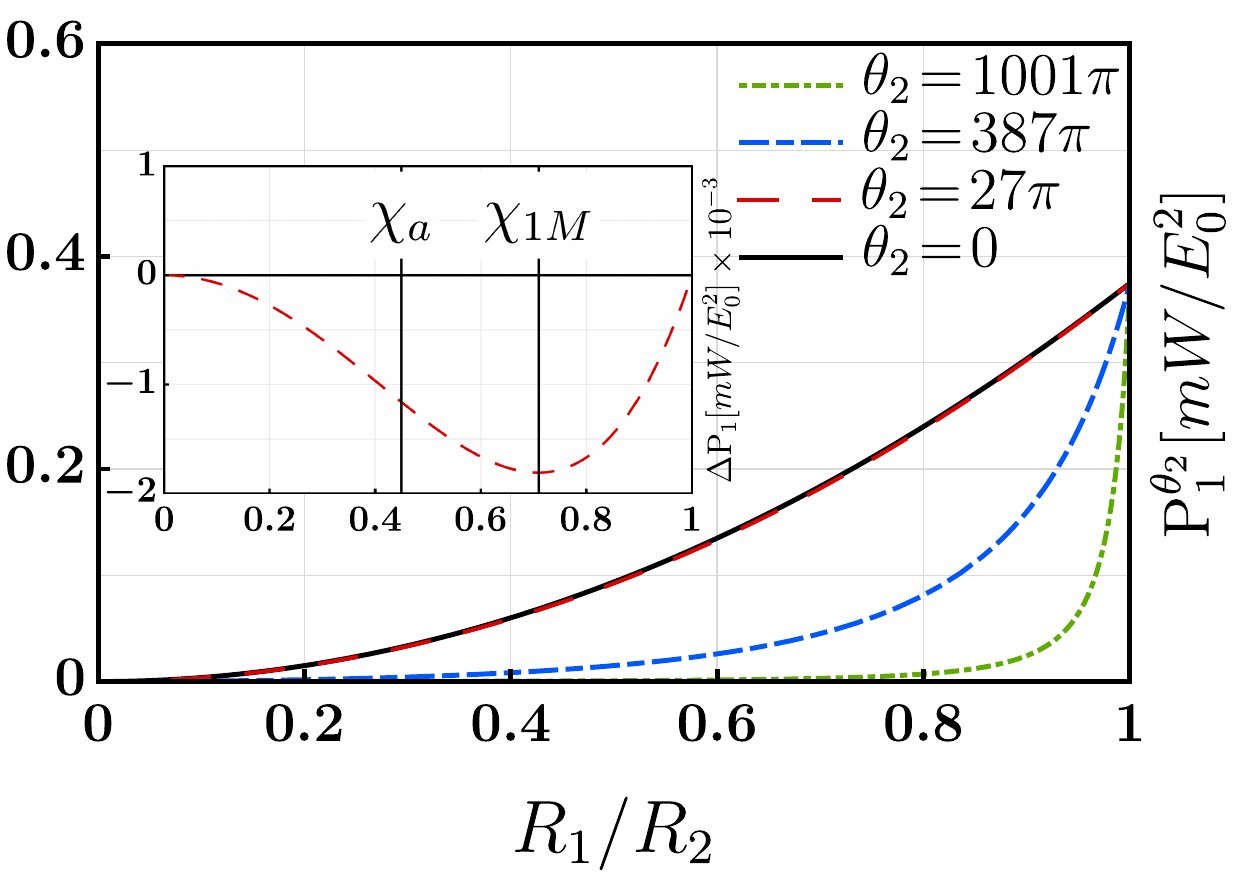}}
\hspace{0.7cm}
  \stackinset{l}{25pt}{t}{12pt}{\Large{(d)}}{\includegraphics[width=0.4\textwidth]{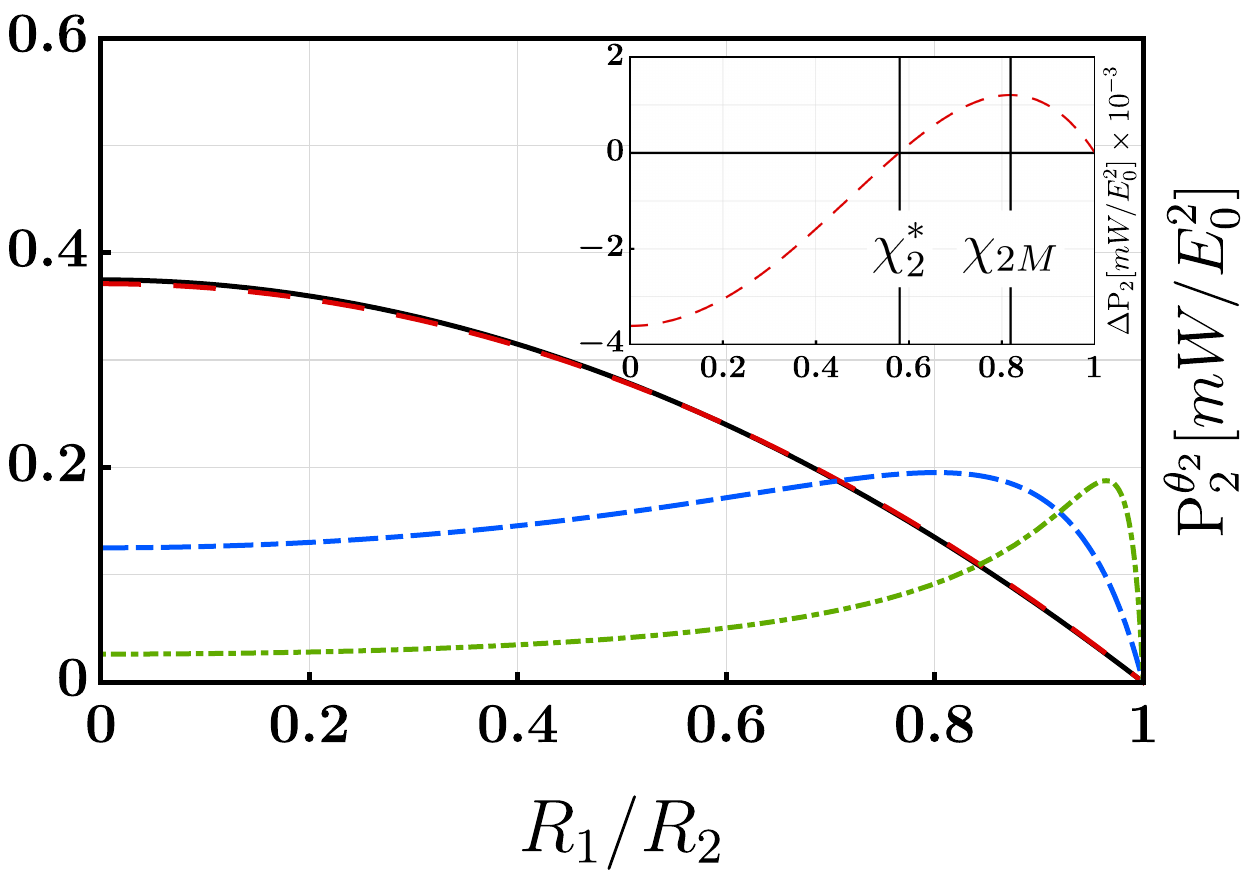}}
\caption{\label{FIG:DensitiyPlotAPwaveguide} In all cases $Z=1$. %
In (a) and (b), $\theta_1=0=\theta_3$ and $\theta_2=27 \pi$ inside the TI.
In (a) $\chi=0.45$ and in (b) $\chi = 0.82$. In (c,d) we show the power transmitted through regions 1 and 2: $P_{1}^{\theta_2}$ and $P_{2}^{\theta_2}$, respectively for $R_2=10 \mu m$. For $\theta_2=0$, the corresponding powers: $P_{1,2}^{\theta_2=0}$, are shown in solid black line. 
For $\theta_2 = 27 \pi$, in the inset of (c) we show $\Delta P_{1}$ . The vertical lines are $\chi_{a} = 0.45$ which defines the geometry of the configuration in (a), and  $\chi_{1M}$ that maximizes the difference. The inset of (d)  shows $\Delta P_{2}$ with the values $\chi_2^\ast$ and $\chi_{2M}$ shown. 
}
\end{figure}

In Fig. (\ref{FIG:DensitiyPlotAPwaveguide}.d), for different values of $\theta_2$, we compare the power transmitted in region 2, $P_{2}^{\theta_2}(\chi)$, to that transmitted in the same region by the background field, $P_2^{\theta_2=0}$  (in solid black line).
Now, we observe that for a given $\theta_2$, there exists a  $\chi=\chi_2^\ast(\theta_2)$ for which the power transmitted in the TI's bulk begins to be greater than  the power transmitted in region 2 ($R_1 \leq \rho \leq R_2$) if the TI were not there, namely: $\Delta P_2 (\chi_2^\ast)>0$.  This occurs when the  $\theta_2\neq 0$ curves cross the solid black line ($\theta_2=0$). Regardless the value of $\theta_2$ such an intersection always occurs, but it is more evident for larger values of $\theta_2$ (compare the red (dashed) curve to the blue (dotted-dashed) or green (smaller dotted-dashed) curves. 

Furthermore, this gain can also be maximized, i.e.,  for that given $\theta_2$, there exists a $\chi = \chi_{2M}(\theta_2)$ such that for $\chi_2^\ast < \chi_{2M} < 1$, the difference $\Delta P_2$ is maximized. In Fig. (\ref{FIG:DensitiyPlotAPwaveguide}.d), we have chosen $\chi = \chi_{2M}(27 \pi)$, precisely to show the maximum gain in region 2, for which:
\begin{equation}
    \textrm{max}({P_{2}^{\theta_2=27 \pi } (\chi)}) = P_{2}^{\theta_2=27 \pi } (\chi_{2M}) =  1.01  \times P_2^{\theta_2=0}.
\end{equation}
This means a 1\% of power gain in region 2  with the TI compared to the case if the TI were not there. The  values $\chi_b=\chi_{2M}$,  $\chi_2^\ast$ and  $\Delta P_2$ and are shown as inset to Fig. (\ref{FIG:DensitiyPlotAPwaveguide}.d).
%
Also, the bigger the $\theta_2$ the largest the gain, however, the closer  $\chi_{2M}$ must be to 1, i.e., higher yields occur for higher $\theta_2$ and through thinner TI sheaths. 

A priori, one could have expected that for a given $\theta_2$ that configuration that minimizes the power transmitted through region 1 (inner vacuum core) is the same configuration that maximizes the power transmitted through region 2 (inside the TI). Namely, the expectation that the power gain through the TI is at the expense of the loss of power in the inner vacuum core, as if the TI sucked power from the inner shells only. Rather surprisingly, this is not the case. In fact, for $\theta_1=0$ we can show that there is no $\chi$ that  minimizes $P^{\theta_2}_1$ and simultaneously maximizes $P^{\theta_2}_2$, implying that both geometry optimization procedures described are in fact independent. To contrast the explanation above, the physical reason for this would then be that not only does the TI confines the EM field in its bulk by depleting the EM field in the inner vacuum, but does so with the field exterior to the TI too. This is why the density plot of the Poynting distribution, for a fixed angular direction and fixed external radius $R < \rho$, is fainter in Fig. (\ref{FIG:DensitiyPlotAPwaveguide}.b) than it does in Fig. (\ref{FIG:DensitiyPlotAPwaveguide}.a).




\section{Summary and outlook}
\label{SEC:conclus}
In this study, we find purely transverse electromagnetic ($\mathcal{T\!E\!M}$) fields propagating parallel to the axis of cylindrical topological insulating (TI) media, that are not possible with topologically trivial materials alone. The media considered were a single TI or several coaxial cylindrical TI layers. These \tem fields propagate both outside the cylindrical TIs as asymptotic free solutions and inside each of the geometries with linear dispersion relation as in a free medium, without cut-off frequencies and without birefringence. Finite discontinuities of the topological magnetoelectric parameter (TMEP) at the interface between each layer, impose boundary conditions that result in a rotation of the polarization plane of the EM field.
This rotation is different from previously reported Faraday or Kerr rotations for TIs, attesting to a new observable signature of the topological magnetoelectric effect (TME). 

In the case of a single $\theta$-layer case, the field exhibits a peculiar asymmetric quadrupolar distribution in the plane perpendicular to the TI. The magnitude of the rotation in the TI core range from  $\approx 0.63$ degrees (11 mrad) to $\approx 5.63$ degrees (98 mrad) for $\theta_{\textrm{TI}}=3 \pi, 27 \pi$ respectively. This well within experimental sensitivity, and as a polarimetric  signal of the TME it is actually 
competitive with respect to other topological magnetoelectric rotations of the plane of polarization of light in TIs \cite{wu_quantized_2016}, even coming close to the enhanced rotation effects reported in \cite{crosse_theory_2017}. These, however, as we have emphasized, are Faraday and/or Kerr rotations and ours, though of similar magnitude, are of a different nature.

For the case of 2 $\theta$-layers, the field  propagates along the TI sheath as in a optical fiber, but in an omnidirectional way. Its confinement  can be controlled varying the geometry ($\chi$) and the value of $\theta$.
We mentioned that in 
$\theta$-ED, the space of solutions enlarges given that Earnshaw's theorem no longer applies %
\footnote{The implications of Earnshaw's theorem for the existence (or not) of $\mathcal{T\!E\!M}$ solutions are more ``stringent'' for cases with conductors. We defer this study in the context of $\theta$-ED for \cite{underprep}.}.
Here we find a non-trivial $\mathcal{T\!E\!M}$ wave solution confined in a topological insulator sheath. This solution is possible due to the modification of the BCs by the topological magnetoelectric properties of TIs. In ordinary Maxwell theory, such a solution is impossible. Hosting $\mathcal{T\!E\!M}$ wave solutions in optical fibers is highly cherished in optics and photonics. Our transverse EM field solutions are dispersion-free and have a linear dispersion relation. This implies that wave packets do not spread during propagation and that there is no cut-off frequency. Additionally, by having no conductors at all, Ohmic losses are reduced. Furthermore,  $\mathcal{T\!E\!M}$ waves propagate in an omnidirectional manner, i.e.,  the EM field propagates inside the TI sheath without undergoing total internal reflection at the TI's walls. This contrasts  TE or TM propagation, in which the field undergoes successive internal reflections and the incident angle cannot exceed the critical one above which the EM field no longer reflects but rather gets refracted outside the fiber. This attribute is highly appealing for miniaturized devices, as it allows the TI-optical fiber to be bent in any angle.

Our results point towards new directions for light manipulation purposes and for studying new manifestations of the TME. Looking ahead, these results could be more appealing by dispensing of the background asymptotic EM field. By adapting the methodology of \cite{medel_electromagnetic_2023}, one could analyze the possibility  to  confine  the $\mathcal{T\!E\!M}$ fields in a finite region.
To disentangle the TME from other optical effects, we have kept $Z=\sqrt{\mu/\epsilon}=1$. However, some of the observables we have found are proportional to $Z \tilde \theta$ or $Z^2 \tilde \theta^2$, therefore one could explore the conditions under which a certain TI could acquire epsilon-near-zero (ENZ) behavior to enhance  the effects.
Lastly, analytical solutions with several TI cylinders are cumbersome. Preliminary numerical calculations indicate that an ad hoc array of several parallel TI cylinders would result in a considerable gain of observable signatures of the TME, due to an enhancement of the Poynting vector by means of superposition. These and other open questions will be dealt with in \cite{underprep}.

\section{Acknowledgements}
\label{SEC:acknow}
We thank L. F. Urrutia and A. Mart\'in-Ruiz for useful comments, and Instituto de Ciencias Nucleares at UNAM, for the hospitality during early stages of this work. Both authors also acknowledge support from the project CONACyT CF/2019/428214. S. F. has been funded by  Scholarship Program/BECAS DOCTORADO UNAB. M. C. has been funded by DGI-UNAB Project DI-16-20/REG.

\bibliography{TEMbib}

\end{document}